\documentclass[%
 reprint,
showpacs,showkeys,
 amsmath,amssymb,
 aps,
 prb,
]{revtex4-1} 
\usepackage{graphicx}
\usepackage{dcolumn}
\usepackage{bm}
\usepackage[breaklinks]{hyperref}
\usepackage{array}
\usepackage[normalem]{ulem}

\renewcommand{\arraystretch}{1.25}

\begin{document}

\preprint{APS/123-QED}

\title{A first-principles study of the electronic structure of Iron-Selenium; Implications for electron-phonon superconductivity}

\author{Alexander P. Koufos}
\author{Dimitrios A. Papaconstantopoulos}
\affiliation{
  School of Physics, Astronomy, and Computational Sciences, George Mason University, Fairfax, Virginia, 22030, USA
}

\author{Michael J. Mehl}
\affiliation{
  Center of Computational Material Sciences, Naval Research Laboratory, Washington, D.C., USA
}
\date{\today}

\begin{abstract}
We have performed density functional theory (DFT) calculations using the linearized augmented plane wave method (LAPW) with the local density approximation (LDA) functional to study the electronic structure of the iron-based superconductor Iron-Selenium (FeSe). In our study, we have performed a comprehensive set of calculations involving structural, atomic, and spin configurations. All calculations were executed using the tetragonal lead-oxide or P4/nmm structure, with various volumes, c/a ratios and internal parameters. Furthermore, we investigated the spin polarization using the LDA functional to assess ferromagnetism in this material. The paramagnetic LDA calculations find the equilibrium configuration of FeSe in the P4/nmm structure to have a volume of 472.5au$^3$ with a c/a ratio of 1.50 and internal parameter of 0.255, with the ferromagnetic having comparable results to the paramagnetic case. In addition, we  calculated total energies for FeSe using a pseudopotential method, and found comparable results to the LAPW calculations. Superconductivity calculations were done using the Gaspari-Gyorffy and the McMillan formalism and found substantial electron-phonon coupling. Under pressure, our calculations show that the superconductivity critical temperature continues to rise, but underestimates the measured values.
\end{abstract}
\pacs{71.15.Mb, 71.20.-b, 74., 74.20.Pq, 74.70.Xa}
\keywords{Iron-Selenium; Electronic Structure; Superconductivity; Density functional Theory; Iron-based Superconductors}

\maketitle
\section*{Introduction} \
Iron-based superconductors are the newest addition to high-temperature superconductivity. Current experimental findings have made many believe that the superconductivity may not be due to electron-phonon interaction\cite{Mizuguchi2008,Medvedev2009,Louca2010}. Spin-fluctuations and spin-density waves have been suggested as mechanisms for the high-temperature superconductivity, but without quantitative assessment. It is therefore important to fully study the electronic structure of these materials, and its implications on superconductivity.

\begin{figure}[!h]\centering
  \includegraphics*[angle=0,width=.4\textwidth]{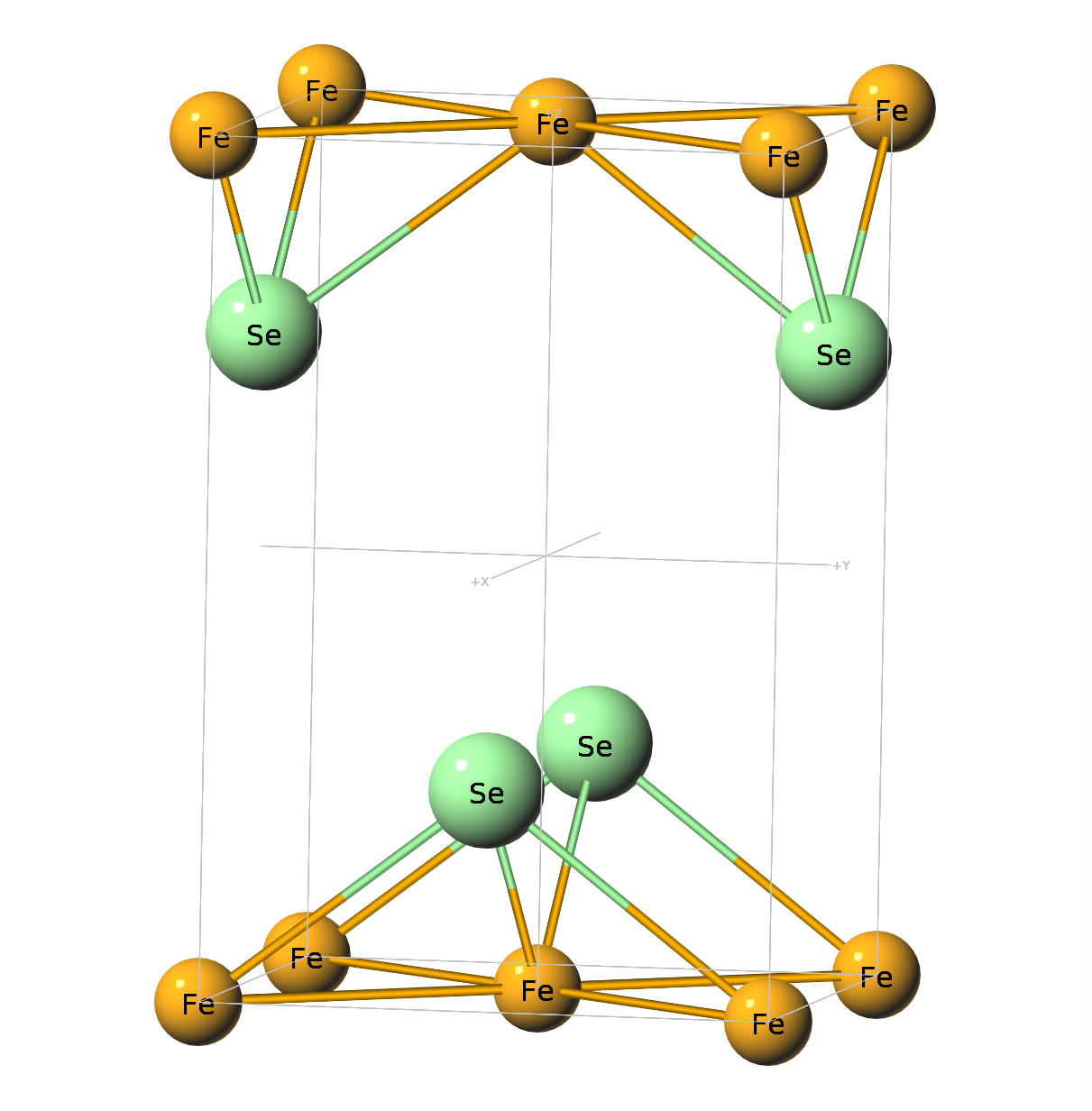}
  \caption{(Color online) Ground state structure of FeSe, {\em
      Strukturbereicht} designation B10.  The space group is
    P4/nmm-D$_{4d}^7$ (\#129).  The iron atoms are on (2a) Wyckoff
    sites, while the selenium atoms are on (2c) sites.}
  \label{fig:B10}
\end{figure}

Iron-selenium (FeSe) has the simplest structure of the current
iron-based superconductors.  As shown in Fig.~\ref{fig:B10},
under ambient conditions\cite{McQueen2009,Medvedev2009,Louca2010,Mizuguchi2008,Diko2012} it forms in the tetragonal PbO
structure, {\em Strukturbericht} B10,\cite{Ewald1931}, space group
P4/nmm-D$_{4d}^7$ (\#129).  The Fe atoms are fixed at the (2a)
Wyckoff position $(000)$, while the Se atoms are at the (2c) Wyckoff
position $(0 1/2 z)$.

There have been several studies of iron-selenium both computationally and experimentally\cite{McQueen2009,Medvedev2009,Louca2010,Mizuguchi2008,Diko2012,Subedi2008,
Singh2010,Bazhirov2012,Kumar2010,Ksenofontov2010,Winiarski2012}. Most of the  computational works use the experimental equilibrium results as input without optimization of all parameters through first-principles. In this work we have performed calculations using the experimental parameters as well as calculations based on first-principles energy minimization.

\begin{figure*}[!ht]\centering
  \includegraphics*[angle=0,width=.8\textwidth]{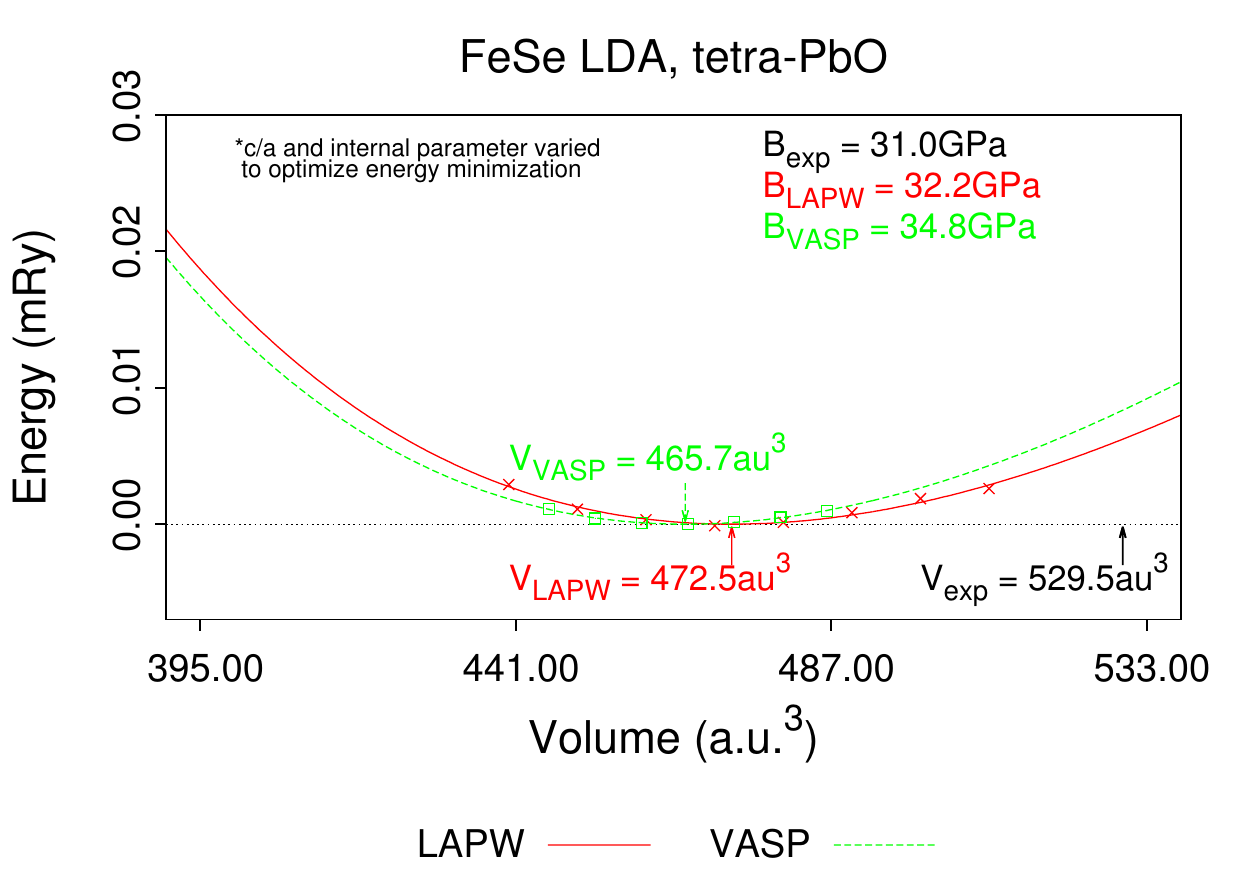}
  \caption{(Color online) Total energy of FeSe using the LAPW with LDA functional and VASP code with LDA functional. Both methods performed energy minimization to obtain the optimal energy for each volume. Although the equilibrium volume is underestimated, the bulk modulus value is in very good agreement with experiment.}
  \label{fig:TE}
\end{figure*}

\section{Computational Details}
\label{sec:Methods}
Most calculations performed in this paper used the Linearized Augmented Plane Wave (LAPW)\cite{singh94}, implementation of Density Functional Theory (DFT)\cite{hohenberg64}. LAPW wave functions were used for the valence band, further augmented by local orbitals for the semi-core states, using a code developed by Krakauer, Wei, and Singh\cite{krakauer:lapw, singh:lapw}. Exchange and correlation effects were approximated by the Hedin-Lundqvist\cite{hedin71} parametrization of the Local Density Approximation (LDA)\cite{kohn65}. The rigid muffin-tin approximation (RMTA) code developed by Papaconstantopoulos and Klein\cite{papa77} was used to apply the Gaspari-Gyorffy theory\cite{Gaspari1972}. A $\Gamma$-centered k-point mesh of 75 points was used for total energy and density of states (DOS) calculations. A larger mesh of 904 k-points was used for the calculations of energy bands. All calculations used a basis set size of 6x6x6, 7 core states (equilibrium state of argon) for iron and 9 core states (argon + 3d states) for selenium resulting in 28 total valence electrons. Local orbitals were also used with energies of 0.308$Ry$ for 3s and 3p iron and -0.548$Ry$ and 0.158$Ry$ for 3s and 3p selenium, respectively. All calculations used fixed muffin-tin radii of 2.0$bohr$ for Fe and Se atoms. Structural optimization was executed via energy minimization with respect to both the tetragonal lattice constants $a$ and $c$ and the internal Selenium parameter $z$. As a check on our LAPW results, we performed calculations on the B10 structure using the Vienna {\em Ab initio} Simulation Package (VASP)\cite{Kresse93:vasp1,Kresse94:vasp2,Kresse93:vasp3} using the VASP implementation\cite{Kresse99:ustopaw} of the Projector Augmented-Wave (PAW) method\cite{Blochl94:paw}. To ensure convergence we used a plane-wave cutoff of 500eV, and used the same k-point mesh as in the LAPW calculations.

\section{Total Energy of Iron-Selenium}
\label{sec:TE}
Fig.~\ref{fig:TE} shows the optimized (variation of volume, c/a and internal parameter z) total energy calculations performed with our LAPW code, as well as our optimized VASP calculations with the LDA functional. Both LAPW and VASP methods with LDA functional underestimate the measured lattice parameters by 4.5\% and 2.2\% for the $a$ and $c$ parameters, respectively. This was expected since the LDA functional usually underestimates the lattice parameters, although for simple materials the difference is usually smaller. Bulk modulus results were overestimated by 3.9\% using the LAPW method. The calculated and experimental structural results and bulk moduli of FeSe are given in Table~\ref{tab:TE}. Structural results from another optimized study of FeSe using PAW and LDA functional completed by Winiarski, et. al.\cite{Winiarski2012} agree with our calculations. They found lattice parameters of $a=6.7902$, $c=10.177au$ and $z=0.257$. Although it is not explicitly stated in their paper, these correspond to underestimations of experimental values of 4.7\% and 2.5\% for $a$ and $c$ parameters, respectively. Comparing our structural results under pressure with the Winiarski\cite{Winiarski2012} paper, we find comparable results for all pressures. 

All calculations presented in this paper are for paramagnetic FeSe. We also performed ferromagnetic calculations which yield nearly the same results for total energy to the paramagnetic calculations. Furthermore, the calculated equilibrium parameters are nearly equivalent to the paramagnetic case, and therefore show no significant difference between the two cases. Further calculations using the ferromagnetic, antiferromagnetic, and other magnetic orders should be considered for further study, but are beyond the scope of this paper.

\begin{table}[!ht]
  \caption{\label{tab:TE}Volume, c/a ratio, internal parameter ($z$) and bulk moduli for FeSe at ambient pressure. Calculated results are from optimized calculations of the corresponding method and LDA functional. Experimental results are taken from Kumar et. al.\cite{Kumar2010} and Ksenofontov, et. at.\cite{Ksenofontov2010}.}
  \begin{ruledtabular}
  \newcolumntype{d}[1]{D{.}{.}{#1}}
  \begin{tabular}{@{}d{3.1}d{1.3}d{1.3}d{1.3}d{2.1}@{}}
    \multicolumn{1}{c}{volume}&\multicolumn{1}{c}{$a$}&\multicolumn{1}{c}{$c/a$}&\multicolumn{1}{c}{$z$}&\multicolumn{1}{c}{Bulk Mod.}\\
    \multicolumn{1}{c}{(au$^3$)}&\multicolumn{1}{c}{(au)}&&\multicolumn{1}{c}{(au)}&\multicolumn{1}{c}{(GPa)}\\ 
    \hline
    \multicolumn{5}{c}{\textbf{LAPW}}\\ 
    472.5&6.804&1.50&0.255&32.2\\ \hline
    \multicolumn{5}{c}{\textbf{VASP}} \\
    465.7&6.771&1.50&0.256&34.8\\ \hline
    \multicolumn{5}{c}{\textbf{Experiment}} \\
    529.5\cite{Kumar2010}&7.121\cite{Kumar2010}&1.465\cite{Kumar2010}&0.269\cite{Kumar2010}&31.0\cite{Ksenofontov2010}\\
  \end{tabular}
  \end{ruledtabular}
\end{table}

\section{Electronic Structure}
\label{sec:DOSandBands}
Density of States (DOS) and energy band calculations were performed using the LAPW results from the optimized LDA total energy calculations. These calculations are performed down to 76\% $V_{exp}$, where $V_{exp}=529.5$au$^3$ from Table~\ref{tab:TE}. This corresponds to pressures as high as 8GPa. The DOS results are then used to calculate superconductivity properties.

\begin{figure}[!h]\centering
  \includegraphics*[angle=0,width=.5\textwidth]{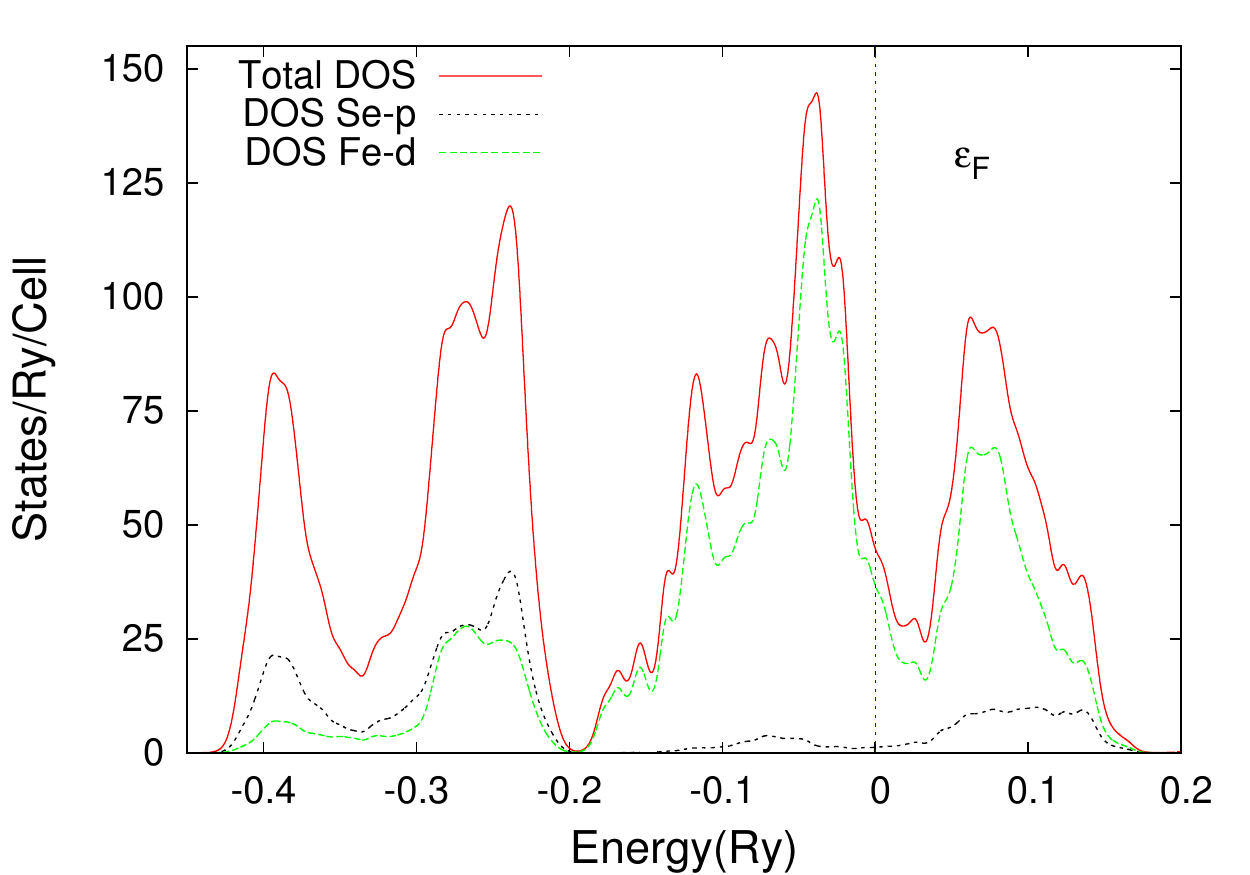}
  \caption{(Color online) Total, Se-p, and Fe-d density of states of FeSe at ambient pressure. Notice that the Fe-d component is the largest contributor to the total DOS around the Fermi level. Dashed vertical (blue) line represents the Fermi level.}
  \label{fig:DOS}
\end{figure}
\begin{figure}[!ht]\centering
  \includegraphics*[angle=0,width=.5\textwidth]{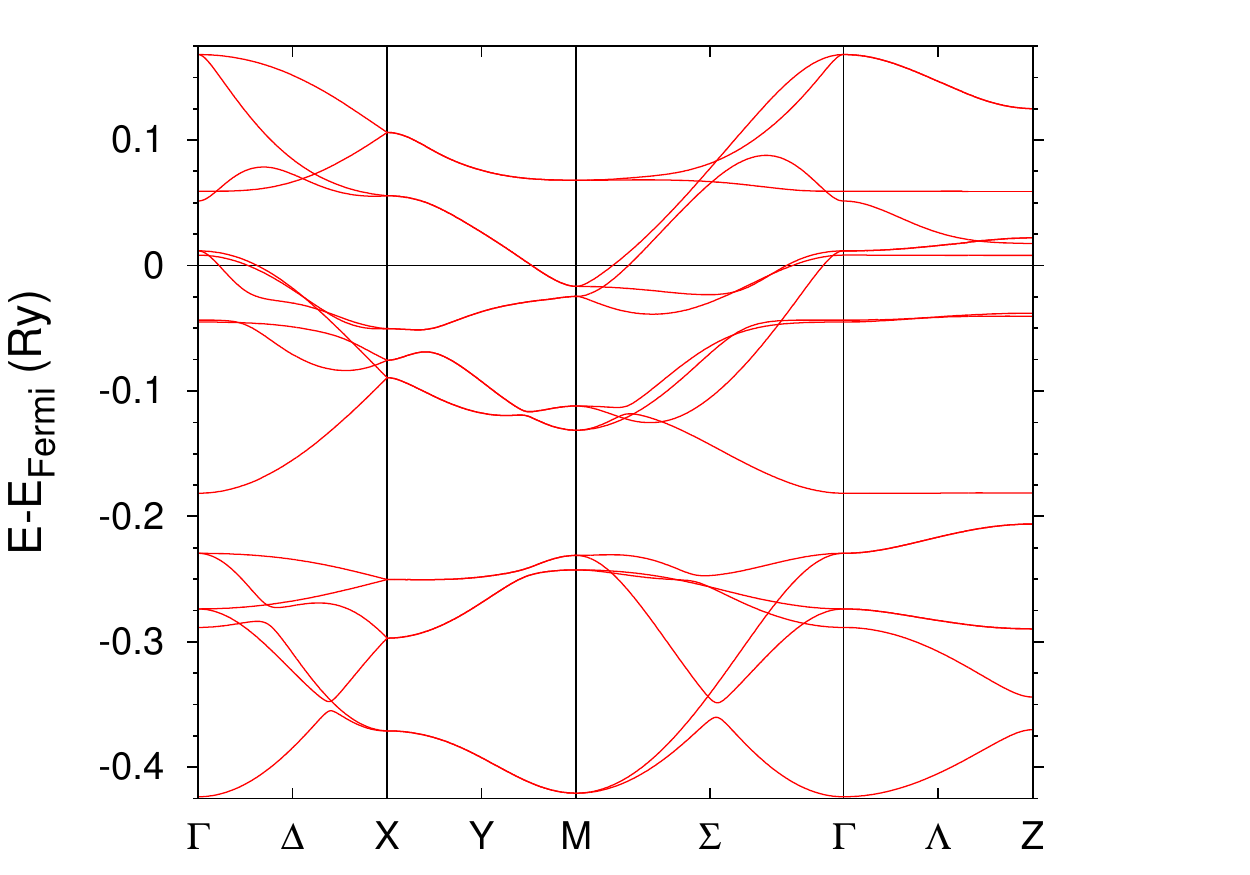}
  \caption{(Color online) Energy bands of FeSe at ambient pressure. Solid horizontal (black) line represents the Fermi level.}
  \label{fig:bands}
\end{figure}

Fig.~\ref{fig:DOS} shows the total, Se-p, and Fe-d decomposed DOS for FeSe at ambient pressure, i.e. $V_{exp}$. The d-component of the Fe DOS is the largest contributor to the total DOS near and at the Fermi level, $E_F$. Around $E_F$, these Fe-d states are localized and do not extend into the interstitial region, while the Se-p and the Fe-d tails have significant contribution in the interstitial space as shown in Fig.~\ref{fig:DOS}. The Se-s semi-core states, not shown in Fig.~\ref{fig:DOS}, are found approximately 1Ry below $E_F$. Fig.~\ref{fig:bands} displays the energy bands of FeSe at ambient pressure. These two figures are in agreement with calculations of the DOS and energy bands performed by other groups\cite{Subedi2008,Singh2010,Bazhirov2012}. Table~\ref{tab:DOS} gives a comparison of the DOS between our results for the experimental lattice parameters with those from Subedi et. al.\cite{Subedi2008}, and Bazhirov and Cohen\cite{Bazhirov2012}. The specific lattice parameters used for the DOS calculations presented in Table~\ref{tab:DOS} are a=7.114a.u., c=10.1154a.u. and z=0.2343 for both Subedi et. al.\cite{Subedi2008} and us. However, it is unclear what structural parameters were used in reference \cite{Bazhirov2012}. DOS calculations for various a, c, and z leads to different values at the Fermi surface. These differences in structural parameters are influencing the calculation of superconducting properties and will be discussed in more detail in the following section. The differences in $N(E_F)$ shown in Table~\ref{tab:DOS} are probably due to the details of the method used to calculate the DOS and possibly the number of k-points.

\begin{table}[!ht]
    \caption{\label{tab:DOS}Comparison of the total DOS at the ambient pressure. Our DOS calculations, as well as those of Subedi et. al.\cite{Subedi2008}, are for the experimental structural parameters with the internal parameter z=0.2343. It is not entirely certain the structural parameters used to calculate the values given by Bazhirov and Cohen\cite{Bazhirov2012}.}
    \begin{ruledtabular}
    \newcolumntype{d}[1]{D{.}{.}{#1}}
    \begin{tabular}{cd{2.1}} 
      &\multicolumn{1}{c}{$N(E_F)$ (states/Ry/cell)}\\ \hline
      This paper&32.3\\
      Subedi\cite{Subedi2008}&26.0\\
      Bazhirov\cite{Bazhirov2012}&21.0\\
    \end{tabular}
    \end{ruledtabular}
\end{table}

\section{Superconductivity}
\label{sec:Supercon}

As mentioned, the DOS calculations are used to calculate superconductivity parameters. For each atom type, we calculate the electron-ion matrix element known as the Hopfield parameter, $\eta_i$\cite{Hopfield1969}, using the following formula\cite{Gaspari1972}, 
\begin{equation}
  \label{eq:eta}
  \eta_i=N(E_F)<I^2>_i
\end{equation}
where $N(E_F)$ is the total DOS per spin at $E_F$ and $<I^2>_i$ is the electron-ion matrix element for each atom type, which is calculated by the Gaspari and Gyorffy theory (GG)\cite{Gaspari1972}. The electron-ion matrix element is given by the following equation,
\begin{equation}
  \label{eq:I2}
  <I^2>_i=\frac{E_F}{\pi^2N^2(E_F)}\sum^2_{l=0}\frac{2(l+1)sin^2(\delta^i_{l+1}-\delta^i_l)N^i_{l+1}N^i_l}{N_{l+1}^{(1),i}N_l^{(1),i}}
\end{equation}
where $N^i_l$ are the per spin angular momenta ($l$) components of the DOS at $E_F$ for atom type $i$, $N_l^{(1),i}$ are the so-called free-scatterer DOS for atom type $i$, and $\delta^i_l$ are scattering phase shifts calculated at the muffin-tin radius and at $E_F$ for atom type $i$. Free-scatterer DOS are calculated by
\begin{equation}
  \label{eq:scatter}
  N_l^{(1)}=\frac{\sqrt{E_F}}{\pi}(2l+1)\int_0^{R_s}r^2u_l^2(r,E_F)dr
\end{equation}
and scattering phase shifts are calculated by
\begin{equation}
  \label{eq:phase}
  \tan\delta_l(R_S,E_F)=\frac{j'_l(kR_S)-j_l(kR_S)L_l(R_S,E_F)}{n'_l(kR_S)-n_l(kR_S)L_l(R_S,E_F)}
\end{equation}
where $R_S$ is the muffin-tin radius, $j_l$ are spherical Bessel functions, $n_l$ are spherical Newnaun functions, $L_l=u'_l/u_l$ is the logarithmic derivative of the radial wavefunction, $u_l$, evaluated at $R_S$ for different energies, $k=\sqrt{E_F}$ and $u_l$ is computed by solving the radial wave equation at each k point in the Brillouin zone. The Hopfield parameter is then used to calculate the electron-phonon coupling constant, which is obtained using the following equation from McMillan's strong-coupling theory\cite{McMillan1968}
\begin{equation}
  \label{eq:lambda}
  \lambda=\sum^2_{i=1}\frac{\eta_i}{M_i<\overline\omega^2>}
\end{equation}
where $M_i$ is the atomic mass of atom type $i$, and $<\overline\omega^2>$ is the average of the squared phonon frequency taken from the experimentally calculated Debye temperature of Ksenofontov, et. at.\cite{Ksenofontov2010}. The Debye temperature is related to the phonon frequency by 
\begin{equation}
  \label{eq:omega2}
  <\overline\omega^2>=\frac{1}{2}\overline\Theta^2_D
\end{equation}
where $\overline\Theta_D$ is found to be 240K\cite{Ksenofontov2010}.
The critical superconductivity temperature is given by the McMillan equation\cite{McMillan1968}, 
\begin{equation}
  \label{eq:tc}
  T_C=\frac{\Theta_D}{1.45}\exp\left[\frac{-1.04(1+\lambda)}{\lambda-\mu^*(1+0.62\lambda)}\right]
\end{equation}
where $\mu^*$ is the Coulomb pseudopotential, given by the Bennemann-Garland equation~\cite{Bennemann1972}
\begin{equation}
  \label{eq:mu}
  \mu^*=0.13\frac{N(E_F)}{1+N(E_F)}
\end{equation}
In equation~\ref{eq:mu}, $N(E_F)$ is expressed in $eV$ and given in a per cell basis. The prefactor 0.13 was chosen such that $\mu^*=0.10$ at the experimental volume. $\Theta_D$ was calculated as a function of volume, V, by the following formula
\begin{equation}
  \label{eq:Theta}
  \Theta_D=C(V-\overline{V})+\overline\Theta_D
\end{equation}
where $C$ is given as the slope between experimental Debye temperature of Ksenofontov, et. at.\cite{Ksenofontov2010} at ambient and 6.9GPa pressures at their corresponding volumes. We have $C=-0.513$, where $\overline{V}$=530.3au$^3$.

In Table~\ref{tab:super}, we show the total DOS at the Fermi level, Hopfield parameters, electron-phonon coupling constants, $\mu^*$ and critical superconductivity temperature for FeSe at various volumes, using our LAPW results. We have included constant c/a and z calculations in Table~\ref{tab:super}, where c/a is set at the experimental value.

\begin{table}[!ht]
    \caption{Total DOS at $E_F$, $N(E_F)$, Hopfield parameters, $\eta$, electron-phonon coupling constants, $\lambda$, $\mu^*$ and critical superconductivity temperature, T$_c$ fo FeSe at various volumes, corresponding to pressures as high a 8GPa. Experimental values are calculated using a fixed c/a and z taken from experimental parameters (c/a=1.4656 and z=0.260/z=0.2343)}
    \begin{ruledtabular}
    \renewcommand{\arraystretch}{1.5}
    \newcolumntype{d}[1]{D{.}{.}{#1}}
    \begin{tabular}{cd{2.2}d{2.2}d{2.2}d{2.3}d{2.3}d{2.3}d{1.4}d{1.1}}
      \multicolumn{1}{c}{V}&\multicolumn{1}{c}{$N(E_F)$}&\multicolumn{1}{c}{$\eta_{Fe}$}&\multicolumn{1}{c}{$\eta_{Se}$}&\multicolumn{1}{c}{$\lambda_{Fe}$}&\multicolumn{1}{c}{$\lambda_{Se}$}&\multicolumn{1}{c}{$\lambda$}&\multicolumn{1}{c}{$\mu^*$}&\multicolumn{1}{c}{T$_c$}\\
      \multicolumn{1}{c}{(au$^3$)}&\multicolumn{1}{c}{($\frac{states}{Ry/cell}$)}&\multicolumn{2}{c}{($eV/\AA^2$)}&&&&&\multicolumn{1}{c}{(K)}\\ \hline
      \multicolumn{9}{c}{\textbf{Experimental c/a; optimized z} (c/a=1.4656; z=0.2343)}\\ \hline
      530&32.28&1.47&0.46&0.500&0.114&0.614&0.10&4.9\\ \hline
      \multicolumn{9}{c}{\textbf{Experimental parameters} (c/a=1.4656; z=0.260)}\\ \hline
      530&44.82&1.50&0.36&0.513&0.090&0.603&0.10&4.6\\ 
      460&37.05&2.15&0.47&0.558&0.088&0.646&0.095&6.8\\
      430&34.23&2.58&0.53&0.600&0.089&0.689&0.093&8.5\\ 
      420&33.44&2.75&0.55&0.618&0.088&0.707&0.0927&9.4\\
    \end{tabular}
    \end{ruledtabular}
    \label{tab:super}
\end{table}
 
First we note that at the experimental volume we obtain T$_c$ $\approx$5K reasonably close to the measured value of 8K. We also note that if the c/a ratio and z are held constant, specifically at the experimental values, throughout the DOS calculations for the given volumes, an increase in all T$_c$ values are found for increased pressure (decreasing volume), as seen in Table~\ref{tab:super}. Although, the total $N(E_F)$ is found to monotonically decrease during the increase in pressure, the parameter $\eta$ undergoes a rapid increase. This decrease in total $N(E_F)$ also contributes to a subtle reduction in the $\mu^*$ at larger pressures. It is also of interest that the change of internal parameter does not seem to influence the overall superconductivity properties. We continue to see the increase of $\eta$ due to the complexity of equation~\ref{eq:I2}, that gives the electron-ion matrix elements, $<I^2>$. This shows that the total DOS at the Fermi level is not the only, nor the major, influence in calculating superconductivity properties. It is important to note from Table~\ref{tab:super} that $\lambda_{Fe}$ is approximately 6 to 7 times larger than $\lambda_{Se}$. This is not surprising since the Fe states dominate near Ef, but also justifies our use of the Debye temperature in estimating the average phonon frequency. Varying the c/a ratio for a given volume does, however, cause a significant change in the superconductivity properties. This makes the search of absolute optimized parameters quite hard and time consuming. We show the trend of increasing superconductivity critical temperature for increasing pressure using the experimental structural parameters, with a value of about 9K, which is too small to account for the measured value of 30K.

At ambient conditions, our electron-phonon coupling constant calculation, $\lambda=0.603$, is consistent with the value calculated by Ksenofontov, et. at.\cite{Ksenofontov2010}, $\lambda=0.65$, by inverting the McMillan equation using the measured $\theta_D$.

Other computational papers\cite{Subedi2008,Bazhirov2012} find values of approximately $\lambda$=0.15 using linear response theory. It is not clear to us what is the source of discrepancy between our calculations based on the Gaspari-Gyorffy theory and the linear response theory-based calculations. It is possible that our use of the relationship between average phonon frequency and the Debye temperature is an oversimplification or on the other hand the Brillouin-zone samplings performed in the linear response codes are not sufficiently converged. It would be helpful for resolving this issue to have separate calculations of $<I^2>$ by the linear response method. In any case, our results presented in the following paragraphs regarding our agreement with the small $\lambda=0.23$ we obtained for LaFeAsO needs to be understood.

In this other iron-based superconducting material, LaFeAsO, electron-phonon coupling constant of approximately 0.2\cite{Boeri2008,Mazin2008} have been reported. We have also performed superconductivity calculations of LaFeAsO using the experimental lattice constants\cite{Papaconstantopoulos2010}. Our calculations of the electron-phonon coupling constant for LaFeAsO is consistent with the results of these groups. LaFeAsO is known to be on the verge of magnetic instability\cite{Boeri2008,Mazin2008}. This suggests that spin-fluctuations are important in this material. 

\begin{table}[!ht]
  \caption{Calculated DOS and superconductivity related results of LaFeAsO using LAPW with LDA functional. Experimental critical temperature result is taken from Takahashi et. al\cite{Takahashi2008}}
  \begin{ruledtabular}
  \renewcommand{\arraystretch}{1.5}
  \newcolumntype{d}[1]{D{.}{.}{#1}}
  \begin{tabular}{d{2.3}d{1.3}d{1.3}d{2.3}d{1.3}d{1.3}d{1.3}}
    \multicolumn{7}{c}{\textbf{LaFeAsO}} \\ \cline{3-5}
    \multicolumn{3}{l}{$N(E_F)$ (states/Ry/cell)}\\ \cline{1-3}
    \multicolumn{3}{d{2.3}}{53.964}\\ \hline
    &\multicolumn{1}{c}{$N_s(E_F)$}&\multicolumn{1}{c}{$N_p(E_F)$}&\multicolumn{1}{c}{$N_d(E_F)$}&\multicolumn{1}{c}{$N_f(E_F)$}&\multicolumn{1}{c}{$\eta$ ($eV/\AA^2$)}&\multicolumn{1}{c}{$\lambda$}\\ \hline
    \multicolumn{1}{l}{Fe}&0.029&0.562&44.052&0.014&0.967&0.19\\
    \multicolumn{1}{l}{As}&0.006&0.826&0.439&0.119&0.186&0.03\\
    \multicolumn{1}{l}{La}&0.006&0.126&0.330&0.278&0.034&0.003\\
    \multicolumn{1}{l}{O}&0.009&0.264&0.042&0.005&0.008&0.006\\ \hline
    \multicolumn{1}{c}{$\Theta_D$(K)}&\multicolumn{1}{c}{$\lambda_{Total}$}&\multicolumn{1}{c}{$\mu*$}&\multicolumn{1}{c}{T$_c$}&\multicolumn{1}{c}{T$_c$ (exp)}\\ \cline{1-5}
    \multicolumn{1}{c}{319}&0.23&0.13&0.0&45.0\\
  \end{tabular}
  \end{ruledtabular}
  \label{tab:FeAs}
\end{table}

Table~\ref{tab:FeAs} shows our calculated LAPW DOS and superconductivity results using the LDA functional and experimental parameters for LaFeAsO. The Hopfield parameter of the Fe atom is seen to be quite low for the LaFeAsO material, in contrast to what we find in FeSe. Similarly, the electron-phonon coupling is much larger for FeSe.

\section{Conclusions}
\label{sec:Con}
In this paper we present calculations of the band structure of FeSe which are in good agreement with other works regarding mechanical and electronic properties of this material. We have also presented calculations of the parameters entering the McMillan equation for the superconducting critical temperature. Our view is that an electron-phonon mechanism can explain superconductivity in FeSe at zero pressure, but it does not give enough of an enhancement to the value of T$_c$ under pressure. Comparing with the multicomponent compound LaFeAsO we see the following picture emerging: In LaFeAsO, a combination of the small value of the calculated parameter $\eta$ and a large value of the measured $\theta_D$ invalidates the electron-phonon coupling. On the other hand, in the case of FeSe, the combination of large $\eta$ and small $\theta_D$ supports electron-phonon coupling, at least at zero pressure. 

\begin{acknowledgments}
Research funded in part by DOE grant DE-FG02-07ER46425 and by ONR grant N00014-09-1-1025. We wish to thank Igor Mazin and Michelle Johannes for useful discussions.
\end{acknowledgments}
\bibliography{Iron-Selenium}{}
\end{document}